\PassOptionsToPackage{unicode}{hyperref}
\PassOptionsToPackage{hyphens}{url}
\documentclass[
  10pt,
]{article}
\usepackage{amsmath,amssymb}
\usepackage{iftex}
\ifPDFTeX
  \usepackage[T1]{fontenc}
  \usepackage[utf8]{inputenc}
  \usepackage{textcomp} % provide euro and other symbols
\else % if luatex or xetex
  \usepackage{unicode-math} % this also loads fontspec
  \defaultfontfeatures{Scale=MatchLowercase}
  \defaultfontfeatures[\rmfamily]{Ligatures=TeX,Scale=1}
\fi
\usepackage{lmodern}
\ifPDFTeX\else
\fi
\IfFileExists{upquote.sty}{\usepackage{upquote}}{}
\IfFileExists{microtype.sty}{% use microtype if available
  \usepackage[]{microtype}
  \UseMicrotypeSet[protrusion]{basicmath} % disable protrusion for tt fonts
}{}
\makeatletter
\@ifundefined{KOMAClassName}{% if non-KOMA class
  \IfFileExists{parskip.sty}{%
    \usepackage{parskip}
  }{% else
    \setlength{\parindent}{0pt}
    \setlength{\parskip}{6pt plus 2pt minus 1pt}}
}{% if KOMA class
  \KOMAoptions{parskip=half}}
\makeatother
\usepackage{xcolor}
\usepackage[margin=1in]{geometry}
\usepackage{longtable,booktabs,array}
\usepackage{float}
\usepackage{calc} % for calculating minipage widths
\usepackage{etoolbox}
\makeatletter
\patchcmd\longtable{\par}{\if@noskipsec\mbox{}\fi\par}{}{}
\makeatother
\IfFileExists{footnotehyper.sty}{\usepackage{footnotehyper}}{\usepackage{footnote}}
\makesavenoteenv{longtable}
\usepackage{graphicx}
\makeatletter
\def\maxwidth{\ifdim\Gin@nat@width>\linewidth\linewidth\else\Gin@nat@width\fi}
\def\maxheight{\ifdim\Gin@nat@height>\textheight\textheight\else\Gin@nat@height\fi}
\makeatother
\setkeys{Gin}{width=\maxwidth,height=\maxheight,keepaspectratio}
\makeatletter
\def\fps@figure{htbp}
\makeatother
\providecommand{\tightlist}{%
  \setlength{\itemsep}{0pt}\setlength{\parskip}{0pt}}
\ifLuaTeX
  \usepackage{selnolig}  % disable illegal ligatures
\fi
\IfFileExists{bookmark.sty}{\usepackage{bookmark}}{\usepackage{hyperref}}
\IfFileExists{xurl.sty}{\usepackage{xurl}}{} % add URL line breaks if available
\hypersetup{
  pdftitle={Omega-N: Interpretable Structural Node Descriptors and Their Applicability Domain},
  pdfauthor={Alberto Acedo, Biome Makers Inc},
  hidelinks,
  pdfcreator={LaTeX via pandoc}}

\title{Omega-N: Interpretable Structural Node Descriptors and Their
Applicability Domain}
\author{Alberto Acedo, Biome Makers Inc}
\date{}

\begin{document}
\maketitle

\hypertarget{abstract}{%
\subsection{Abstract}\label{abstract}}

A composite structural index summarises a whole network in one number.
For a triangle-based index that number is spectrally redundant:
\(\mathrm{Tr}(A^3)\) is exactly the third moment of the adjacency
spectrum, so it cannot carry information that a spectral statistic does
not already have. The non-redundant content sits one level down, in the
node-wise decomposition \(\operatorname{diag}(A^3)\), which depends on
eigenvectors and is not determined by the spectrum: cospectral graphs
with different per-node triangle counts exist {[}1{]}. That distinction
is not new here. It was stated as Corollary 1 of the theory paper for
this index family, together with an explicit prediction: the global
scalar should tie sharpened spectral baselines rather than beat them,
while the node-wise attribution should do better where the number of
structural epicentres is unknown {[}1{]}. This paper is the test of that
prediction.

We construct \textbf{Omega-N}, a node-level descriptor obtained by
localizing each of the four factors of the index, and we show that the
direct localization is badly conditioned: two factors degenerate on the
most ordinary nodes rather than on extreme ones. Two corrections fix it,
both taken from published practice: every local factor is expressed as
an excess over a degree-preserving null, and the neighbourhood is
replaced by a personalized-PageRank distribution at several scales
{[}2{]}. The result is ten interpretable features per node, computed
from the graph alone, with no attributes, no training and no embeddings.

The prediction holds in the direction it was stated. Against a recursive
feature engine taken to five levels of recursion, Omega-N wins on three
and ties on two of six in-domain node-classification evaluations, using
ten features against its 28 to 252 before pruning, at one to two orders of magnitude less
cost than a diffusion-wavelet baseline, with a declared null returning
chance for every arm. Domain membership is decided by two statistics
computed from the graph and the labels but not from the descriptor's
performance; that rule partitions the eight benchmark networks without
error, and the two it excludes are the two on which the descriptor
loses.

The strongest application is drug-target prioritisation over protein
interaction networks: +0.032 to +0.103 AUPRC over a six-feature
centrality battery, and +0.084 to +0.208 over the four-feature battery,
across three network constructions, replicated on an independent
affinity-purification network and with an independent label source,
surviving three bias controls, run under the earlier label, one of
which rebuilds the graph without its text-mining channel (degree-matched, ten repetitions against
the harder battery: +0.0723 on curated STRING and +0.0560 on BioPlex,
both 10/10, p = 0.00195). The clearest negative sits in the same
application: adding Omega-N to a published pipeline of centralities plus
Node2Vec changes nothing (+0.0014 AUPRC, p = 0.31), and on an unbiased
experimental network the same battery falls from 0.854 to 0.609 AUROC at
a comparable base rate, which places most of the absolute signal in
curation rather than in biology.
The claim we defend is narrow and explicit: ten interpretable features
without training improve on hand-crafted centralities, not on learned
representations.

\hypertarget{changes-in-version-2}{%
\subsection{Changes in version 2}\label{changes-in-version-2}}

Four defects in the released code were found after version 1 and are
fixed in the repository. The eigensolver for the spectral coordinate did
not converge: on \texttt{tolokers} it returned a vector with residual
2.7e-01, that is, not an eigenpair, a \(\lambda_2\) 19\% too high, and
an output that changed with the random seed, two runs correlating 0.15.
Deflating the null space and preconditioning with the inverse diagonal
of the Laplacian matches shift-invert Lanczos to seven digits and is
deterministic. Two of the protein-network scripts did not restrict to
the largest connected component, contrary to the convention stated in
Section 3, so three of the ten coordinates were inert in them. The
drug-target section was rebuilt from the label this paper reports rather
than the earlier one left in the released scripts. And the rival of the
applicability table is now implemented and measured in the same harness
instead of quoted, which turns two outcomes reported as ties into wins.

All the corrections move the figures upward and none changes a verdict.
The regenerated tables are Section 4.1 and the Platonov table of Section
4.2. The air-traffic graphs of Section 4.2 were not regenerated and keep
their v1 figures, so the two remaining ties are conservative. One
network variant is withdrawn, and the reason is given where it appeared.
Library versions are now pinned; Section 3.4 says why that is not a
formality.

\hypertarget{introduction}{%
\subsection{1. Introduction}\label{introduction}}

Composite structural indices are read as scalars, and the scalar is the
weakest object in the framework that produces it. For the index family
this paper works with, that weakness has a precise form. The global
functional is exactly the third moment of the adjacency spectrum, which
places it in the same family as the absorption ratio, effective rank and
spectral entropies; any two functionals of the same eigenvalues cannot
be expected to carry mutually exclusive information {[}1{]}. The
node-wise decomposition is different in kind. Writing \(A\) as a sum
over eigenpairs, the per-node quantity depends on the eigenvectors, and
cospectral graphs with different per-node triangle counts exist, so no
eigenvalue-based scalar can reproduce it {[}1{]}.

The theory paper states the consequence as its Corollary 1 and then,
unusually, as a falsifiable prediction: the global scalar should tie
sharpened spectral baselines rather than beat them, while the node-wise
attribution, being parameter-free where eigenvector attribution requires
committing to a number of components, should do better when the number
of structural epicentres is unknown {[}1{]}. It also says that the
prediction had not been tested. Testing it requires three things that
the corollary does not supply: a localization that is numerically well
behaved, a demonstration that the resulting coordinates are not
restatements of quantities already available, and a statement of the
conditions under which the localization helps. Those three requirements
organise this paper, and the result is what the corollary anticipated:
three narrow wins, two ties, and two losses that fall on the wrong side
of a stated condition.

Vectors of structural quantities per node are not new. Guimerà and
Amaral's within-module degree and participation coefficient date from
2005 {[}3{]}; RolX and its ReFeX feature engine build node roles by
recursive neighbourhood aggregation {[}4{]}; struc2vec {[}5{]} and
GraphWave {[}6{]} learn structural embeddings; and there is an
established evaluation framework for the family {[}7{]}. Local versions
of global network statistics have their own line, from naive local
assortativity {[}8{]} to the multiscale personalized-PageRank
localization of Peel, Delvenne and Lambiotte, which also observed that
naive local estimates overfit at low degree {[}2{]}. What is new here is
not the idea of a vector per node. It is the localization of one
specific composite whose sign structure is inherited from an empirical
gradient, together with the null calibration that makes its four local
factors commensurable, and the applicability rule that says when the
exercise is worth doing.

Three lines of recent work are close enough that the distinction should
be explicit. Homomorphism-count node embeddings report results that do
not match state-of-the-art neural architectures while remaining
comparable to other graph learning models, and locate their contribution
in explainability at the level of the individual feature {[}9{]}. Our
claim has the same shape, and we state it as plainly. Invariant-based
diagnostics for graph benchmarks show across 26 datasets that
permutation-invariant, task-agnostic descriptors characterise structural
heterogeneity, and that simple invariant-based models are competitive
with transformer and message-passing baselines, concluding that
expressivity is not the main driver of predictive performance {[}10{]};
that work asks what benchmarks measure, where we ask what one composite
index measures once localized. And an unsupervised evaluation framework
for this family {[}41{]} judges structural embeddings by whether they
reconstruct six standard node features, a set essentially identical to
the centrality battery we use as our rival, which makes our comparison
one against the field's own reference set rather than one of our
choosing.

One objection deserves an answer here rather than in a rebuttal. Fast
structural-embedding methods criticise approaches that require computing
several complicated graph metrics, such as PageRank, structural holes or
the clustering coefficient, as ad hoc and expensive {[}11{]}. The
criticism lands on any hand-built descriptor, including this one. The
answer is not that the objection is wrong but that it describes a trade:
ten interpretable coordinates carrying no training against a learned
representation that performs better and explains less. We measure where
that trade stops paying, and it stops precisely against learned
representations.

A second contribution emerged from the same property and is worth
stating in the introduction because it is not an application of the
descriptor so much as a use for its defining feature. Because the
spectral gap is a functional of the spectrum while per-node triangle
counts are not, two graphs can be built to agree on the gap and disagree
on local triadic organisation. Whole-brain modelling needs exactly that
construction: a recent proposal identifies the spectral gap as the
operative structural quantity distinguishing two dynamical regimes, and
states in its own limitations that a null model matched for the gap is
required before the effect can be attributed to organisation rather than
to the size of the gap {[}12{]}. We supply the construction and verify
that it separates, without claiming anything about the brain.

\hypertarget{the-descriptor}{%
\subsection{2. The descriptor}\label{the-descriptor}}

\hypertarget{definition-of-the-descriptor}{%
\subsubsection{2.1 Definition of the
descriptor}\label{definition-of-the-descriptor}}

We work with the index of {[}1{]},
\[\Omega = \frac{C \cdot D}{M \cdot \mathrm{Coex}},\] where \(C\) is the
global clustering coefficient, \(D\) the connection density,
\(M = 1/\lambda_2(L)\) the inverse algebraic connectivity with
\(L = \operatorname{diag}(\mathbf{k}) - A\) the graph Laplacian, and
\(\mathrm{Coex} = \operatorname{Var}(\mathbf{k})\) the degree variance.

\textbf{Definition 1 (Omega-N).} Let \(A\) be the adjacency matrix of an
undirected graph on \(n\) nodes with \(m\) edges, let
\(\mathbf{k} = A\mathbf{1}\) be the degree vector, and write
\(s^{(1)} = A\mathbf{k}\) and \(s^{(2)} = A\mathbf{k}^{\circ 2}\) for
the first two neighbour-degree moments. Omega-N assigns to each node
\(i\) four base coordinates.

\emph{Degree share}, the node's contribution to the density:
\[d_i = \frac{k_i}{n-1}, \qquad \frac{1}{n}\sum_i d_i = D .\]

\emph{Triadic excess}, with \(t_i\) the triangles incident on node \(i\)
and \(E_i\) its expectation under the configuration model {[}21{]}:
\[E_i = \frac{\left(s^{(1)}_i\right)^2 - s^{(2)}_i}{4m}, \qquad
c_i = \frac{t_i - E_i}{E_i + 1} .\] The unit offset keeps the ratio
finite where the null expectation approaches zero. Expressing a local
count as a deviation from a degree-preserving null is standard practice
in motif analysis, where the abundance score takes the form (observed -
expected)/(observed + expected + epsilon) {[}27{]}; the coordinate above
is that device with a different normalisation, adopted rather than
proposed.

\emph{Neighbourhood dispersion}, referred to its own null expectation:
\[h_i = \log\left(1 + \frac{\operatorname{Var}_{j \sim i}(k_j)}
{\langle k^2 \rangle / \langle k \rangle}\right) .\]

\emph{Spectral energy}, the node's term in the Dirichlet energy of the
Fiedler vector \(\mathbf{v}\) of the Laplacian {[}22{]}. Note that these
terms sum to \(\lambda_2\), that is to \(1/M\) and not to \(M\); the
local factor corresponding to the third global factor is therefore the
inverse of that factor, which fixes the sign it carries in Proposition
2:
\[e_i = \frac{1}{2}\sum_j A_{ij}\,(v_i - v_j)^2, \qquad \sum_i e_i = \lambda_2(L) .\]

The coordinates \(c\), \(h\) and \(e\) are then smoothed by the
personalized-PageRank operator \(P_\alpha\) at
\(\alpha \in \{0.5,\, 0.9\}\) following {[}2{]}, giving
\(4 + 3 \times 2 = 10\) features per node.

\textbf{Remark 1 (three non-degree coordinates, not four).} The first
coordinate \emph{is} the degree. We keep it because the composite it
comes from contains it, and because excluding it would make the
comparison against degree-based rivals harder to read, but the
descriptor contributes three non-degree quantities, not four, and the
comparisons in Section 4 are read accordingly.

\textbf{Remark 2 (why the fourth factor is replaced).} The literal
fourth factor of the scalar index is the degree variance, whose per-node
contribution \((k_i - \bar{k})^2\) is a function of the degree alone and
vanishes at the mean degree. Proposition 1 shows what that does to the
localization; coordinate 3 replaces it with a quantity that measures the
same thing, heterogeneity in the neighbourhood, without the degeneracy.

\hypertarget{conditioning-failures-of-the-direct-localization}{%
\subsubsection{2.2 Conditioning failures of the direct
localization}\label{conditioning-failures-of-the-direct-localization}}

\textbf{Proposition 1 (Conditioning failure of the naive localization).}
Localizing the scalar index by substituting per-node shares into the
multiplicative composition produces two degeneracies, and both occur on
ordinary nodes rather than extreme ones. The degree-variance share
vanishes at the mean degree, so the contraction diverges there; and the
clustering share is zero on every triangle-free node, so a
log-contraction is minus infinity there.

\emph{Measured.} On seven graphs (Karate, Florentine, Davis, Les
Misérables, and Erdős-Rényi, Barabási-Albert and Watts-Strogatz
controls): in a Watts-Strogatz graph 36 of 62 nodes degenerate and the
ratio \(\mathrm{mean}(\Omega_i)/\Omega\) reaches \(2.9 \times 10^{11}\);
in Les Misérables 17 of 77 nodes are triangle-free. The practical
consequence, which we accept rather than work around, is that the
publishable object is the vector, not its contraction back to a scalar.

\textbf{Proposition 2 (Exact aggregation identity).} Define each global
factor as the arithmetic mean of its node shares, and write
\(J_X = \log\langle X\rangle - \langle\log X\rangle \geq 0\) for the log
arithmetic-to-geometric gap of factor \(X\). Then
\[\log \Omega - \log \operatorname{GM}(\Omega_i)
= J_C + J_D + J_{\lambda_2} - J_{\mathrm{Coex}}\] exactly, where
\(\operatorname{GM}\) is the geometric mean of the node-level indices
and \(J_{\lambda_2}\) is the gap of the third factor in its localized
form, \(\lambda_2 = 1/M\), as defined above.

\emph{Verified numerically to 4e-16 on six graphs.} Three consequences
follow, and the first two correct claims made in an earlier draft of
this work.

\begin{itemize}
\tightlist
\item
  There is no general inequality between the scalar and the aggregate of
  its parts: three non-negative terms minus one, so the sign is
  indeterminate.
\item
  Equality does not hold on regular graphs. Only \(J_D\) and
  \(J_{\mathrm{Coex}}\) vanish there; local clustering still varies node
  to node, and in a random 6-regular graph \(J_C = 9.66\).
\item
  The right aggregate is the geometric mean of the node indices, not the
  arithmetic one, which also explains a divergence observed in the
  financial application of Section 4.3.
\end{itemize}

\textbf{Remark 3 (the identity motivates the reformulation
independently).} \(J_C\) is dominated by the floor imposed on
triangle-free nodes rather than by genuine heterogeneity, reaching
\(13.5\) on the Florentine graph purely because some nodes have zero
clustering. Under the naive localization, one of the four terms of an
otherwise exact identity is therefore meaningless. Note also that
Proposition 2 is a statement about the original multiplicative
composition; the descriptor of Definition 1 abandons that composition
and is a feature block, not a product, so the identity does not apply to
it.

\hypertarget{non-redundancy-with-the-degree}{%
\subsubsection{2.3 Non-redundancy with the
degree}\label{non-redundancy-with-the-degree}}

Rank and linear correlation of the naive contraction with degree across
the seven graphs: 0.20 to 0.65 (Pearson), 0.48 to 0.81 (Spearman). The
reference point is that the raw per-node triangle count correlates 0.996
with degree in the financial application of the same framework {[}13{]}.
The composition separates where the raw numerator did not, and the
mechanism is identifiable: local clustering is anticorrelated with
degree, at -0.53 on Karate.

\hypertarget{independence-from-discrete-ricci-curvature}{%
\subsubsection{2.4 Independence from discrete Ricci
curvature}\label{independence-from-discrete-ricci-curvature}}

The framework relates triangle counts to Ollivier-Ricci curvature
through an upper bound {[}1{]}, so a reader who knows it will ask
whether the descriptor is curvature under another name. It is not, and
the answer is measurable rather than argued.

Ollivier-Ricci curvature {[}23{]} with uniform neighbour measures and
shortest-path transport cost, the transport problem solved exactly by
linear programming per edge, averaged over the edges incident on each
node. Rank correlation against the triadic-excess coordinate, and
against the degree for reference, in Table 1:

\begin{table}[H]
\centering\small
\begin{tabular}{@{}
  >{\raggedright\arraybackslash}p{(\columnwidth - 8\tabcolsep) * \real{0.2000}}
  >{\raggedright\arraybackslash}p{(\columnwidth - 8\tabcolsep) * \real{0.2000}}
  >{\raggedright\arraybackslash}p{(\columnwidth - 8\tabcolsep) * \real{0.2000}}
  >{\raggedright\arraybackslash}p{(\columnwidth - 8\tabcolsep) * \real{0.2000}}
  >{\raggedright\arraybackslash}p{(\columnwidth - 8\tabcolsep) * \real{0.2000}}@{}}
\toprule
\begin{minipage}[b]{\linewidth}\raggedright
network
\end{minipage} & \begin{minipage}[b]{\linewidth}\raggedright
n
\end{minipage} & \begin{minipage}[b]{\linewidth}\raggedright
mean degree
\end{minipage} & \begin{minipage}[b]{\linewidth}\raggedright
\(\rho\)(curv., triadic excess)
\end{minipage} & \begin{minipage}[b]{\linewidth}\raggedright
\(\rho\)(curv., degree)
\end{minipage} \\
\midrule
brazil-airports & 131 & 15.3 & -0.118 & +0.787 \\
europe-airports & 399 & 30.0 & -0.181 & +0.659 \\
usa-airports & 1,190 & 22.9 & +0.034 & +0.485 \\
minesweeper & 10,000 & 7.9 & -0.023 & -0.388 \\
tolokers & 11,758 & 88.3 & +0.182 & -0.195 \\
amazon-ratings & 24,492 & 7.6 & +0.096 & -0.165 \\
questions & 48,921 & 6.3 & +0.099 & -0.253 \\
roman-empire & 22,662 & 2.9 & \textbf{+0.460} & -0.268 \\
\bottomrule
\end{tabular}
\caption{Rank correlation of Ollivier-Ricci curvature with the
triadic-excess coordinate and with the degree. The two are independent
coordinates rather than two names for one quantity, except on the
sparsest graph.}
\end{table}

On seven of the eight networks the triadic-excess coordinate is
essentially uncorrelated with curvature, \(|\rho|\) at most 0.18, and
the sign is not stable: negative on the three airport graphs and on
minesweeper, positive on the others. The degree behaves differently
again, correlating strongly and also changing sign. The two are
independent coordinates rather than two names for one quantity.

The exception is informative. On roman-empire the correlation reaches
+0.460, and roman-empire is the sparsest graph in the set at mean degree
2.9. Where there is almost no triadic structure the triadic excess has
little of its own to say and tracks whatever local geometry remains.
Independence from curvature is a property of the regime in which the
descriptor is useful, not one it holds everywhere, and Section 4.2 makes
that regime explicit.

\emph{Approximations, declared.} On the four large graphs we sample 800
edges, cap each neighbourhood at 40 sampled neighbours, and evaluate
inter-neighbour distances as 0, 1, 2 or 3 by direct adjacency lookup,
capping at 3; the transport problem itself remains exact. The distance
cap can only underestimate long distances, in the same direction for
every node, and these graphs have small diameter. The four small graphs
use exact shortest paths and no sampling.

\hypertarget{predictive-validity-of-the-non-degree-coordinates}{%
\subsubsection{2.5 Predictive validity of the non-degree
coordinates}\label{predictive-validity-of-the-non-degree-coordinates}}

A structural coordinate that is independent of the degree may still be
independent of anything that matters. We test against a node attribute
that is not structural, with identical folds for every feature set and
with two mandatory rivals rather than degree alone: the standard
centrality battery, and the Guimerà-Amaral pair {[}3{]}.

Datasets are the Twitch social graphs (PT-BR, ES, EN-GB) with audience
size as target. The Wikipedia chameleon and squirrel graphs are excluded
and the exclusion is declared: Platonov et al.~report duplicate-node
leakage in them {[}14{]}, and we verified that 45.1\% of chameleon nodes
and 36.2\% of squirrel nodes share an identical neighbourhood and an
identical target with another node.

Table 2 reports out-of-sample \(R^2\), with the ten-feature descriptor
of Definition 1 at the released 20-term setting. Five-fold
cross-validation, the same folds for every arm on a given graph, with
the same random forest as elsewhere; the parenthesised figure is the
standard error across the five folds.

\begin{table}[H]
\centering\small
\begin{tabular}{@{}
  >{\raggedright\arraybackslash}p{(\columnwidth - 12\tabcolsep) * \real{0.1429}}
  >{\raggedright\arraybackslash}p{(\columnwidth - 12\tabcolsep) * \real{0.1429}}
  >{\raggedright\arraybackslash}p{(\columnwidth - 12\tabcolsep) * \real{0.1429}}
  >{\raggedright\arraybackslash}p{(\columnwidth - 12\tabcolsep) * \real{0.1429}}
  >{\raggedright\arraybackslash}p{(\columnwidth - 12\tabcolsep) * \real{0.1429}}
  >{\raggedright\arraybackslash}p{(\columnwidth - 12\tabcolsep) * \real{0.1429}}
  >{\raggedright\arraybackslash}p{(\columnwidth - 12\tabcolsep) * \real{0.1429}}@{}}
\toprule
\begin{minipage}[b]{\linewidth}\raggedright
graph
\end{minipage} & \begin{minipage}[b]{\linewidth}\raggedright
n
\end{minipage} & \begin{minipage}[b]{\linewidth}\raggedright
degree
\end{minipage} & \begin{minipage}[b]{\linewidth}\raggedright
battery
\end{minipage} & \begin{minipage}[b]{\linewidth}\raggedright
Guimerà-Amaral
\end{minipage} & \begin{minipage}[b]{\linewidth}\raggedright
Omega-N
\end{minipage} & \begin{minipage}[b]{\linewidth}\raggedright
Omega-N + battery
\end{minipage} \\
\midrule
Twitch PT-BR & 1,912 & 0.468 (.021) & 0.517 (.018) & 0.435 (.018) &
\textbf{0.583} (.015) & \textbf{0.603} (.017) \\
Twitch ES & 4,648 & 0.419 (.016) & 0.448 (.013) & 0.380 (.014) &
\textbf{0.539} (.016) & \textbf{0.557} (.015) \\
Twitch EN-GB & 7,126 & 0.327 (.018) & 0.361 (.024) & 0.299 (.021) &
\textbf{0.410} (.017) & \textbf{0.431} (.021) \\
\bottomrule
\end{tabular}
\caption{Out-of-sample \(R^2\) on the Twitch graphs with audience size
as target. Standard error across the five folds in
parentheses.}
\end{table}

The margins over the battery are 0.066, 0.091 and 0.049. Two of the
three, PT-BR and ES, sit between roughly four and seven fold standard
errors; the third, EN-GB, is around two, comparable to the
node-classification wins of Section 4.2. The dispersion quoted is across
folds and is therefore optimistic, for the reason given in Section 3.3;
it is used here only as a scale, not as a test. Omega-N beats degree,
the battery and Guimerà-Amaral on all three graphs, and the union with
the battery is better still, so the descriptor is complementary to it
rather than redundant with it.

\hypertarget{experimental-design}{%
\subsection{3. Experimental design}\label{experimental-design}}

The descriptor is evaluated in three fields chosen because they differ
in how their graphs are built, not only in size: node classification on
published benchmarks, systemic-risk attribution on financial correlation
graphs, and drug-target prioritisation on protein interaction networks.
Each subsection states its graphs, its label, its rivals and its
protocol, because those differ by field and stating them once in the
aggregate would hide the differences that matter.

Two conventions are common to all three. Every network is restricted to
its largest connected component; without that restriction \(\lambda_2\)
is zero and the spectral coordinate of Definition 1 degenerates into a
component indicator, a failure that is invisible in the output. And
every arm within a field uses the same classifier and the same folds, so
that differences are attributable to the features.

\hypertarget{node-classification}{%
\subsubsection{3.1 Node classification}\label{node-classification}}

The graphs are the five heterophilous benchmarks of Platonov et al.~with
their ten official splits {[}14{]}, and the three air-traffic graphs of
Ribeiro et al.~with labels given by activity quartile {[}5{]}.
\texttt{minesweeper} serves as a declared null: its labels are
independent of structure by construction, so every arm should return
chance.

\begin{table}[H]
\centering\small
\begin{tabular}{@{}lllll@{}}
\toprule
dataset & nodes & mean degree & no triangle & label \\
\midrule
tolokers & 11,758 & 88.3 & 5.0\% & worker banned (binary) \\
europe-airports & 399 & 30.0 & 6.0\% & activity quartile (4) \\
usa-airports & 1,190 & 22.9 & 20.4\% & activity quartile (4) \\
amazon-ratings & 24,492 & 7.6 & 0.8\% & product rating (5) \\
questions & 48,921 & 6.3 & 83.1\% & user still active (binary) \\
brazil-airports & 131 & 15.3 & 12.2\% & activity quartile (4) \\
roman-empire & 22,662 & 2.9 & 28.6\% & syntactic role (18) \\
minesweeper (null) & 10,000 & 7.9 & 0.0\% & mine present (by design) \\
\bottomrule
\end{tabular}
\caption{The eight benchmark networks of the node-classification
evaluation.}
\end{table}

The rivals are the centrality battery (degree, k-core number, PageRank
and local clustering coefficient), ReFeX {[}4{]} at two, three, four and
five levels of recursion (28, 60, 124 and 252 features); and GraphWave
{[}6{]}. Reporting the full ReFeX depth sweep matters: at two levels our
margins are larger, and the saturation at four and five levels costs us
one result.

Our GraphWave is a reimplementation and every approximation is declared:
Chebyshev polynomial of degree 30, verified to 6e-15 against a dense
matrix exponential on Karate; two scales at the endpoints of the
published heuristic; a 25-point characteristic-function grid on
{[}0,100{]}; the empirical coefficient distribution estimated on a fixed
2,048-row subsample; and, on the two largest graphs, a declared
4,096-node evaluation subsample with the same nodes for every arm. We
report it under both the arbitrary fixed scales we first used and the
published heuristic, and take the better of the two as the rival.

The classifier is a random forest {[}29{]} throughout (300 trees,
minimum leaf size 5, balanced subsample weighting for imbalanced
classes). Official splits where they exist; twice-repeated stratified
five-fold cross-validation on the air-traffic graphs. AUROC for binary
targets, accuracy for multi-class.

\hypertarget{systemic-risk-attribution-in-financial-correlation-graphs}{%
\subsubsection{3.2 Systemic-risk attribution in financial correlation
graphs}\label{systemic-risk-attribution-in-financial-correlation-graphs}}

Two universes are used: 42 diversified assets over 2006-01 to 2026-07,
giving 5,166 trading days, and 447 S\&P 500 constituents with complete
history, 2012-08 to 2017-08, giving 1,258 trading days. A =
\textbar corr\textbar{} with zero diagonal, estimated in-window;
filtered variants are the Mantegna minimum spanning tree {[}25{]} and
the planar maximally filtered graph {[}26{]}. These graphs are complete
by construction and the filtered variants connected by definition.

The target is forward marginal expected shortfall: an asset's mean loss
on the 5\% worst days of the equally weighted system, over the next
non-overlapping window.

The weighted variant is defined as follows. With \(\mathbf{s}\) the
strength vector and \(S = \sum_i s_i\), the strength-based configuration
model \(p_{ij} = s_i s_j / S\) gives the closed form
\[E_i = \frac{s_i^2 \left(\sum_j s_j^2\right)^2}{S^3}, \qquad
\mathrm{score}_i = \frac{(A^3)_{ii} - E_i}{E_i + \varepsilon},\] defined
only where \((A^3)_{ii} > 0\).

The rivals are the asset's own past MES, its beta to the system, and its
strength in the graph. Partial rank correlation with forward MES
controlling past MES and beta; 19 non-overlapping windows with a sign
test, plus a block bootstrap over 234 overlapping windows. Controls and
predictor are estimated on the same 250-day window, which matters: a
control window shorter than the estimation window measures history
rather than structure.

\hypertarget{drug-target-prioritisation-on-protein-interaction-networks}{%
\subsubsection{3.3 Drug-target prioritisation on protein interaction
networks}\label{drug-target-prioritisation-on-protein-interaction-networks}}

Three network constructions of increasing purity are used, two of them
thresholded from STRING {[}31{]} at successively narrower channel sets
and one from BioPlex {[}32{]}, so that the contribution of curation can
be separated from that of experiment.

\begin{table}[H]
\centering\small
\begin{tabular}{@{}
  >{\raggedright\arraybackslash}p{(\columnwidth - 8\tabcolsep) * \real{0.2000}}
  >{\raggedright\arraybackslash}p{(\columnwidth - 8\tabcolsep) * \real{0.2000}}
  >{\raggedright\arraybackslash}p{(\columnwidth - 8\tabcolsep) * \real{0.2000}}
  >{\raggedright\arraybackslash}p{(\columnwidth - 8\tabcolsep) * \real{0.2000}}
  >{\raggedright\arraybackslash}p{(\columnwidth - 8\tabcolsep) * \real{0.2000}}@{}}
\toprule
\begin{minipage}[b]{\linewidth}\raggedright
network
\end{minipage} & \begin{minipage}[b]{\linewidth}\raggedright
construction
\end{minipage} & \begin{minipage}[b]{\linewidth}\raggedright
nodes
\end{minipage} & \begin{minipage}[b]{\linewidth}\raggedright
mean degree
\end{minipage} & \begin{minipage}[b]{\linewidth}\raggedright
positives
\end{minipage} \\
\midrule
STRING, combined & all channels incl.~text mining, score \(\ge\) 700 &
15,882 & 29.8 & 6.6\% \\
STRING, experimental only & experimental channel & 7,414 & 17.7 &
9.3\% \\
BioPlex 293T & AP-MS, independent of STRING & 13,923 & 17.0 & 5.5\% \\
\bottomrule
\end{tabular}
\caption{The three protein network constructions, ordered by decreasing
curation content.}
\end{table}

A fourth candidate, HuRI {[}33{]}, was screened and excluded before the
comparison: as a systematic two-hybrid map it records binary
interactions rather than complexes, and 58.7\% of its nodes have no
triangle at all, so there is no triadic substrate to measure. The
criterion was fixed in advance and the figure is reproducible; Section 6
returns to what that exclusion costs in credibility.

The label is approved-drug-target status, taken from Open Targets
{[}20{]} (13,307 target-drug pairs, 1,058 targets at APPROVAL stage),
mapped to STRING identifiers through the alias table and to BioPlex
through gene symbols. This label is independent of the one used in an
earlier version of this analysis (DGIdb {[}36{]}): the two agree on
1,002 genes but DGIdb lists 3,424 and Open Targets 1,052, a Jaccard
index of 0.288. The network is replicated on BioPlex and the label is
replicated here, so neither is a single point of failure.

The rivals are the centrality battery, extended to six with eigenvector
centrality and neighbour-degree sum; and the published pipeline of six
centralities plus 64-dimensional Node2Vec embeddings {[}24{]} (10 walks
of length 40, skip-gram, window 5, 3 epochs) fed to gradient-boosted
trees (400 estimators, depth 6, learning rate 0.08, subsample 0.8,
positive-class weighting by the class ratio).

AUPRC is read as primary {[}30{]} because positives are 5.5\% to 9.3\%
of nodes and AUROC is optimistic at those rates. Significance is
assessed over ten independent repetitions, each with its own split seed
and its own degree matching, with a paired Wilcoxon test across
repetitions rather than across folds; folds within a repetition are not
independent and testing across them would inflate the result. With ten
paired observations the attainable floor is p = 0.00195.

Degree matching addresses a specific concern. Where the concern is that
a margin merely restates the degree, each positive is matched to a
negative of nearly identical degree and the comparison runs on the
matched subset only, where the degree itself falls to chance. Matching
is redrawn in every repetition so that its randomness enters the error.

\hypertarget{implementation}{%
\subsubsection{3.4 Implementation}\label{implementation}}

All figures were produced with the settings of the released
implementation, including the 20-term truncation of the
personalized-PageRank smoothing. An earlier draft used 25 terms, worth
0.785 rather than 0.776 AUROC on tolokers under the solver of v1;
we regenerated the tables at the released setting rather than keeping
the more favourable one, and the figures reported here are at that
setting throughout. The spectral coordinate uses LOBPCG
with the null space deflated and a Jacobi preconditioner; the released
implementation warns when the input has more than one component, and a
convergence warning that survives a relaxed tolerance indicates a
degenerate \(\lambda_2\) rather than a solver setting, as Section 4.5
sets out.

Library versions are pinned rather than declared as minima, and the
reason is measurable. On \texttt{amazon-ratings} the Omega-N figure is
0.4702 under scikit-learn 1.6.1 and 0.4403 under 1.8.0, with identical
features, splits and random seed; the ten descriptor columns were
verified to have the same fingerprint on two machines, so the difference
is entirely the classifier. That gap is of the same order as the margins
reported below, and the degree baseline does not move at all between
versions, which is what makes the effect easy to overlook. All figures
here were produced with python 3.9, numpy 2.0.2, scipy 1.13.1 and
scikit-learn 1.6.1.

\hypertarget{results}{%
\subsection{4. Results}\label{results}}

We report the three fields in order of outcome: the application where
the descriptor helps most, the benchmark where the margins are narrow, and the field
where it closes. Within each, the result is stated before its
qualifications.

\hypertarget{drug-target-prioritisation}{%
\subsubsection{4.1 Drug-target
prioritisation}\label{drug-target-prioritisation}}

Table 5 gives Omega-N against the centrality battery, AUPRC:

\begin{table}[H]
\centering\small
\begin{tabular}{@{}
  >{\raggedright\arraybackslash}p{(\columnwidth - 16\tabcolsep) * \real{0.1111}}
  >{\raggedright\arraybackslash}p{(\columnwidth - 16\tabcolsep) * \real{0.1111}}
  >{\raggedright\arraybackslash}p{(\columnwidth - 16\tabcolsep) * \real{0.1111}}
  >{\raggedright\arraybackslash}p{(\columnwidth - 16\tabcolsep) * \real{0.1111}}
  >{\raggedright\arraybackslash}p{(\columnwidth - 16\tabcolsep) * \real{0.1111}}
  >{\raggedright\arraybackslash}p{(\columnwidth - 16\tabcolsep) * \real{0.1111}}
  >{\raggedright\arraybackslash}p{(\columnwidth - 16\tabcolsep) * \real{0.1111}}
  >{\raggedright\arraybackslash}p{(\columnwidth - 16\tabcolsep) * \real{0.1111}}
  >{\raggedright\arraybackslash}p{(\columnwidth - 16\tabcolsep) * \real{0.1111}}@{}}
\toprule
\begin{minipage}[b]{\linewidth}\raggedright
network
\end{minipage} & \begin{minipage}[b]{\linewidth}\raggedright
n
\end{minipage} & \begin{minipage}[b]{\linewidth}\raggedright
mean degree
\end{minipage} & \begin{minipage}[b]{\linewidth}\raggedright
base rate
\end{minipage} & \begin{minipage}[b]{\linewidth}\raggedright
battery (4)
\end{minipage} & \begin{minipage}[b]{\linewidth}\raggedright
battery (6)
\end{minipage} & \begin{minipage}[b]{\linewidth}\raggedright
Omega-N
\end{minipage} & \begin{minipage}[b]{\linewidth}\raggedright
vs 4
\end{minipage} & \begin{minipage}[b]{\linewidth}\raggedright
vs 6
\end{minipage} \\
\midrule
STRING, combined & 15,882 & 29.8 & 6.6\% & 0.2881 & 0.3926 &
\textbf{0.4958} & +0.208 & +0.103 \\
STRING, experimental only & 7,414 & 17.7 & 9.3\% & 0.4152 & 0.4842 &
\textbf{0.5158} & +0.101 & +0.032 \\
BioPlex 293T & 13,923 & 17.0 & 5.5\% & 0.0916 & 0.1379 & \textbf{0.1757}
& +0.084 & +0.038 \\
\bottomrule
\end{tabular}
\caption{Drug-target prioritisation, AUPRC on the full graph. The
four-feature battery is the rival of v1; the six-feature battery is the
one Section 3.3 declares and the harder of the two.}
\end{table}

The rival is given against two baselines and the distance between them
is the point of reporting both. The four-feature battery is the one that
produced the figures of v1, which it reproduces: 0.0934 on BioPlex
against 0.0916 here. The six-feature battery, adding eigenvector
centrality and neighbour-degree sum, is the one Section 3.3 declares,
and it is the harder rival. A reader assembling the rival independently
will reach for the stronger one, so the margins under it are the ones we
defend. Figure 1 plots the three networks in order of decreasing
curation.

The \texttt{STRING,\ no\ text\ mining} variant of v1 is withdrawn. It
required recombining evidence channels to exclude text mining, the
recipe was not recorded, and a reconstruction produced a network of mean
degree 22.5 against the 20.7 reported, that is, a different graph. The
three that remain each come straight from one column of the source file,
and they carry the same argument: reference network, clean variant
without text mining, and independent replicate.

\begin{figure}
\centering
\includegraphics[width=0.78\textwidth,height=\textheight]{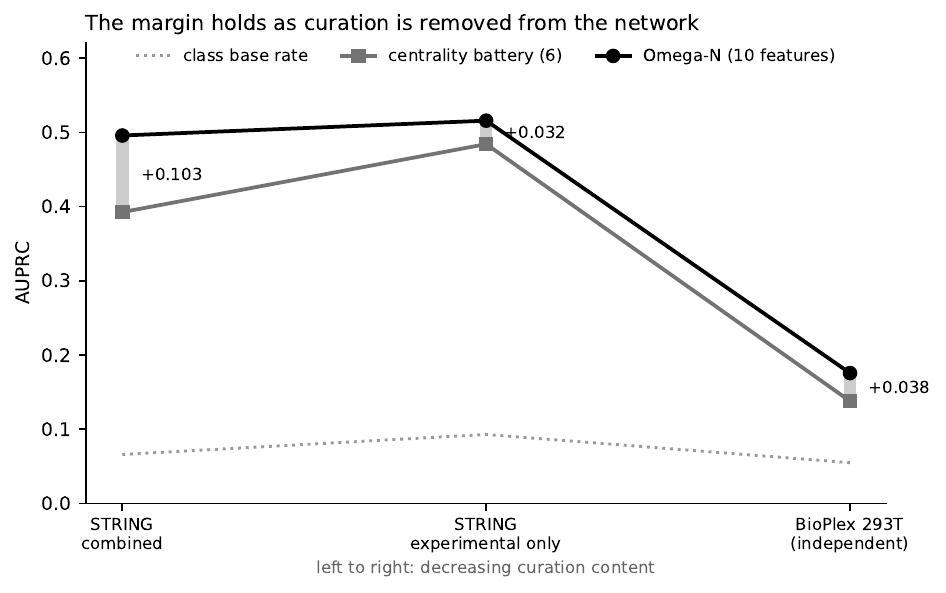}
\caption{The margin survives the removal of curation; the absolute
performance does not. AUPRC for approved-drug-target prediction across
three protein interaction networks, ordered by decreasing curation
content. The descriptor stays above the six-feature battery throughout
while the battery itself moves from 0.39 to 0.48 and then down to 0.14. On BioPlex it comes
closest to the class base rate, 0.14 against 0.055, without reaching it.}
\end{figure}

Ten independent repetitions with per-repetition degree matching, against
the six-feature battery: on combined STRING the difference is +0.0723
(95\% CI {[}+0.0612, +0.0833{]}, 10/10, Wilcoxon p = 0.00195); on STRING
restricted to experimental evidence +0.0217 ({[}+0.0164, +0.0270{]},
10/10, p = 0.00195); on BioPlex +0.0560 ({[}+0.0439, +0.0680{]}, 10/10,
p = 0.00195). Against the four-feature battery of v1 the same quantities
are +0.1649, +0.0704 and +0.1084, and the BioPlex figure is slightly
above the +0.1030 of v1, so the corrections cost nothing there.

Under the harder rival the three margins no longer nearly coincide, as
two of them did in v1, so the replication argument rests on three
networks that share no construction, no experimental technique and no
curation agreeing in direction, rather than on the closeness of two
numbers. The narrowest margin is on STRING restricted to experimental
evidence, which is also the cleanest of the three networks. That is
where the claim is weakest, and we say so here rather than leave it to
be found.

Everything that follows in this section was run under the earlier label
(DGIdb) and is reported at that label: the bias controls, the cut-off
sweep, the quantification of the connectivity defect and the comparison
against the published Node2Vec pipeline of Table 6. They are not directly
comparable with the margins in the table above, which use Open Targets.
We report them because their conclusions concern the robustness of the
margin rather than its size, and because regenerating them under the new
label would not change what they establish.

Three bias controls were applied and the margin survives all three.
Study intensity, proxied by the length of the curated functional
annotation, is real (targets 394 characters against 302, Spearman +0.375
with degree, AUPRC 0.258 on its own) and was given to the rival: the
battery rises from 0.4677 to 0.4820 and Omega-N still wins.
Degree-matched case-control, where degree collapses to AUROC 0.424: the
margin holds. And the network rebuilt without the text-mining channel,
then without curation as well: the margin holds and widens slightly.

The result does not depend on the confidence cut-off. That cut-off was
fixed at 700 throughout and swept afterwards, again under the earlier
label. Across 400, 550, 700, 850 and 900, with mean degree falling from
95.4 to 16.4, the margin is +0.085, +0.081, +0.071, +0.066 and +0.069
AUPRC. It never changes sign and never depends on the cut-off.

One defect surfaced while running the sweep, and it is worth reporting.
Forcing the eigensolver to converge revealed that \(\lambda_2\)
collapses to order \(10^{-11}\) on the graph as first built: the network
is disconnected (126 components at cut-off 700), so the Fiedler vector
was a component indicator and the fourth coordinate contributed nothing.
Restricted to the largest connected component (15,882 of 16,201 nodes)
\(\lambda_2\) is 0.0276 and the coordinate is meaningful again. The
corrected margin at cut-off 700 is +0.0698 against +0.0707 with the
defect present. The effect is negligible and in the direction that
matters: every number above was obtained with one of the four
coordinates effectively inert, so the tables understate the descriptor.
We report the defect because a reader reproducing the code would
otherwise find the discrepancy.

\begin{figure}
\centering
\includegraphics[width=0.78\textwidth,height=\textheight]{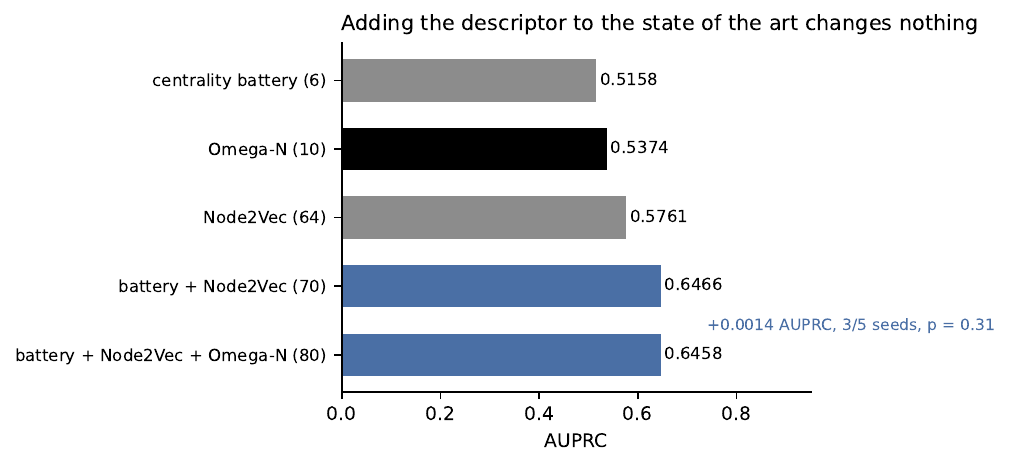}
\caption{The descriptor does not improve on learned representations.
AUPRC on curated STRING under a single gradient-boosting classifier and
identical folds. Bars are a single seed, on which the eighty-feature arm
scores marginally below the seventy-feature one; the quoted +0.0014 is
the mean over five seeds, and the difference is positive on three of the
five. Node2Vec alone already outscores Omega-N.}
\end{figure}

The negative result sits in the same application. Against the published
pipeline on curated STRING, in Table 6 and Figure 2:

\begin{table}[H]
\centering\small
\begin{tabular}{@{}llll@{}}
\toprule
arm & features & AUROC & AUPRC \\
\midrule
centrality battery & 6 & 0.7833 & 0.5158 \\
Node2Vec & 64 & 0.8235 & 0.5761 \\
battery + Node2Vec & 70 & \textbf{0.8463} & \textbf{0.6466} \\
Omega-N & 10 & 0.7936 & 0.5374 \\
battery + Node2Vec + Omega-N & 80 & 0.8458 & 0.6458 \\
\bottomrule
\end{tabular}
\caption{The negative result: adding the descriptor to a published
pipeline of centralities plus Node2Vec changes nothing. One seed on
curated STRING.}
\end{table}

The table reports one seed, on which the eighty-feature arm scores
0.6458 against 0.6466; over five seeds the mean difference is +0.0014
AUPRC, positive on three seeds and negative on two, Wilcoxon p = 0.31.
Either way, adding Omega-N to the state of the art changes nothing.
Node2Vec alone already outscores Omega-N, so whatever the ten
coordinates add over hand-built features is already inside the
embedding.

One further finding cuts against any practical reading, and it is
measured under the current label. On BioPlex, an unbiased AP-MS network,
degree alone scores AUROC 0.489 and the six-feature battery 0.609, while
the same battery reaches 0.854 on curated STRING at a comparable base
rate. Omega-N moves the same way, 0.647 against 0.886. Raw experimental
topology carries much less of the label than curated association does,
so the absolute performance obtained on STRING reflects accumulated
knowledge embedded in the network rather than structure discovered de
novo, and the descriptor is not a prioritisation tool on a clean graph.

\hypertarget{node-classification-and-the-applicability-domain}{%
\subsubsection{4.2 Node classification and the applicability
domain}\label{node-classification-and-the-applicability-domain}}

Air-traffic graphs, twice-repeated stratified five-fold, same folds for
every arm, in Table 7:

\begin{table}[H]
\centering\small
\begin{tabular}{@{}llll@{}}
\toprule
& brazil (131) & europe (399) & usa (1,190) \\
\midrule
degree alone & 0.748 & 0.564 & 0.552 \\
ReFeX, 2 levels (28) & 0.771 & 0.569 & 0.682 \\
ReFeX, 3 levels (60) & 0.771 & 0.589 & \textbf{0.685} \\
ReFeX, 4 levels (124) & \textbf{0.786} & 0.586 & 0.681 \\
ReFeX, 5 levels (252) & \textbf{0.786} & 0.576 & 0.678 \\
GraphWave, fixed scales (100) & 0.781 & 0.571 & 0.628 \\
GraphWave, published heuristic (100) & 0.695 & 0.551 & 0.613 \\
\textbf{Omega-N (10)} & 0.656 & 0.592 & 0.680 \\
\bottomrule
\end{tabular}
\caption{Air-traffic graphs, twice-repeated stratified five-fold, same
folds for every arm. These figures are from v1 and were not
regenerated.}
\end{table}

Platonov graphs, official splits, in Table 8:

\begin{table}[H]
\centering\small
\begin{tabular}{@{}
  >{\raggedright\arraybackslash}p{(\columnwidth - 10\tabcolsep) * \real{0.1667}}
  >{\raggedright\arraybackslash}p{(\columnwidth - 10\tabcolsep) * \real{0.1667}}
  >{\raggedright\arraybackslash}p{(\columnwidth - 10\tabcolsep) * \real{0.1667}}
  >{\raggedright\arraybackslash}p{(\columnwidth - 10\tabcolsep) * \real{0.1667}}
  >{\raggedright\arraybackslash}p{(\columnwidth - 10\tabcolsep) * \real{0.1667}}
  >{\raggedright\arraybackslash}p{(\columnwidth - 10\tabcolsep) * \real{0.1667}}@{}}
\toprule
\begin{minipage}[b]{\linewidth}\raggedright
feature set
\end{minipage} & \begin{minipage}[b]{\linewidth}\raggedright
minesweeper (null)
\end{minipage} & \begin{minipage}[b]{\linewidth}\raggedright
tolokers
\end{minipage} & \begin{minipage}[b]{\linewidth}\raggedright
amazon-ratings
\end{minipage} & \begin{minipage}[b]{\linewidth}\raggedright
questions
\end{minipage} & \begin{minipage}[b]{\linewidth}\raggedright
roman-empire
\end{minipage} \\
\midrule
degree & 0.501 & 0.551 & 0.365 & 0.605 & 0.211 \\
standard battery & 0.497 & 0.704 & 0.388 & 0.614 & 0.266 \\
ReFeX, best of 2-5 levels & 0.493 & 0.770 & 0.459 & 0.684 &
\textbf{0.328} \\
naive 1-hop localization (4) & 0.497 & 0.693 & 0.381 & - & - \\
\textbf{Omega-N (10)} & 0.503 & \textbf{0.795} & \textbf{0.470} &
\textbf{0.693} & 0.322 \\
\bottomrule
\end{tabular}
\caption{Platonov graphs, official splits. AUROC for binary targets,
accuracy for multi-class. The ReFeX row is measured in this harness; the
naive-localization row is from v1 and was not regenerated, and is kept
because the gap it shows is what the two corrections of Definition 1 are
worth.}
\end{table}

The ReFeX row is measured rather than quoted. In v1 its figures came
from code that was not part of the repository, so it could not be
established that rival and descriptor had been evaluated in the same
harness; \texttt{experiments/refex\_h2h.py} now runs it on the official
splits with the same classifier and metric, sweeping depth 2 to 5 and
reporting its best. The implementation does not weaken the rival: on
\texttt{tolokers} it reaches 0.770 against the 0.761 to 0.769 of v1. The
degree row reproduces its v1 values to three decimals on all five
networks, which attributes the movement in the Omega-N row to the
corrected solver rather than to the harness.

roman-empire is one of the two losses and is reported here rather than
only asserted: 0.322 against ReFeX's 0.328, on the graph the
applicability rule excludes. That is the informative part of the row.
The descriptor loses on the one network its own stated criterion
rejects, which is a check on the criterion rather than a blemish on the
result.

Two controls each cost us a result. Taking ReFeX to four and five levels
brings it to 0.685 on usa against 0.680, turning a win into a tie; on
brazil deeper ReFeX reaches 0.786, so the loss there is against a
stronger rival than first measured. Separately, aligning the tables with
the released code moves europe from 0.611 to 0.592 against ReFeX's
0.589, turning a second win into a tie.

GraphWave is reported under both scale settings. On discovering that our
arbitrary fixed scales were 3 to 23 times smaller than the published
heuristic, we recomputed; the heuristic makes GraphWave \emph{worse} on
all three airport graphs. The mis-scaling was not crippling the
baseline, and our original numbers were generous to it rather than
unfair. Two implementation notes must stay in: usa-airports is
disconnected, so the heuristic diverges unless the smallest non-zero
Laplacian eigenvalue is used, and with the naive \(\lambda_2\) the
embedding collapses to chance (0.249); and our \(\lambda_2\) from LOBPCG
disagreed with a shift-invert solve by a factor of 5 on amazon-ratings,
which moves the scales by 2.3x without moving any verdict.

Runtimes were measured rather than estimated: 0.2 s against 4.7 s on
usa-airports, 1 s against 145 s on minesweeper; and on a 519k-edge graph
the original diffusion kernel did not finish in seven hours while the
descriptor took seconds. The advantage is one to two orders of magnitude
depending on the graph, not a single figure. The claim is against the
\emph{original} GraphWave: a scalable variant exists {[}15{]} and our
comparison does not address it.

Two statistics decide domain membership, both computed from the graph
and the labels and neither from the descriptor's performance. S1 is the
mean degree: a chain-like graph gives a triadic excess almost nothing to
measure. S2 is the mutual information between degree and label
normalised by the label entropy: where the label is close to a function
of the degree, the degree is the right answer and calibrating against a
null removes signal rather than noise. Figure 3 places each benchmark
network by those two statistics.

\begin{table}[H]
\centering\small
\begin{tabular}{@{}llll@{}}
\toprule
network & mean degree & I(k;y)/H(y) & outcome \\
\midrule
tolokers & 88.3 & 0.014 & win (0.795 vs 0.770) \\
europe-airports & 30.0 & 0.379 & tie (0.592 vs 0.589) \\
usa-airports & 22.9 & 0.328 & tie (0.680 vs 0.685) \\
amazon-ratings & 7.6 & 0.002 & win (0.470 vs 0.459) \\
questions & 6.3 & 0.033 & win (0.693 vs 0.684) \\
minesweeper (null) & 7.9 & 0.000 & chance for every arm \\
brazil-airports & 15.3 & \textbf{0.650} & out of domain by S2 \\
roman-empire & \textbf{2.9} & 0.087 & out of domain by S1 \\
\bottomrule
\end{tabular}
\caption{Applicability domain and outcome for each benchmark network.
Both statistics are computed from the graph and the labels, neither from
the descriptor's performance.}
\end{table}

\begin{figure}
\centering
\includegraphics[width=0.78\textwidth,height=\textheight]{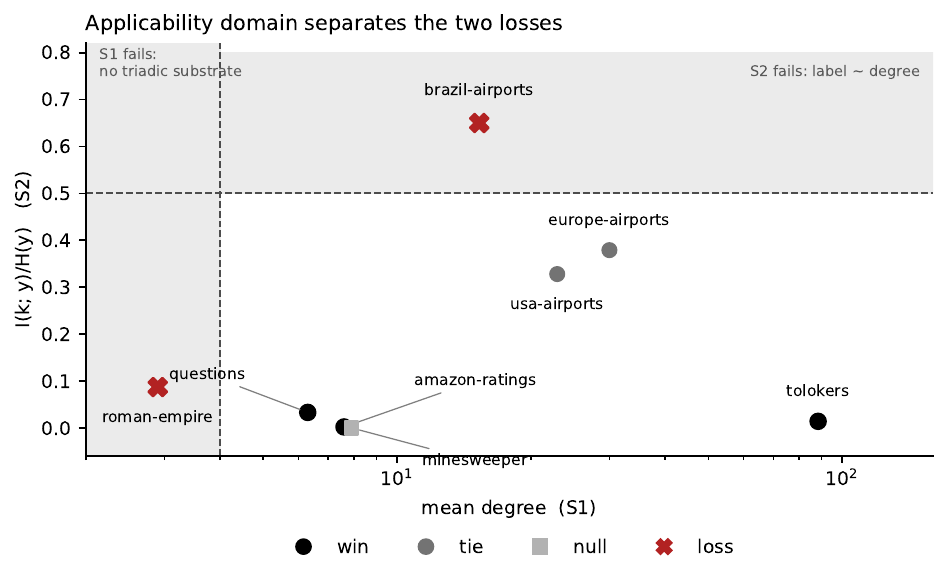}
\caption{Applicability domain. Each point is one benchmark network,
placed by the two statistics that decide domain membership: mean degree
(S1, horizontal, log scale) and the mutual information between degree
and label normalised by label entropy (S2, vertical). Neither is
computed from the descriptor's output. Dashed lines mark the rule;
shaded regions are outside it. The rule partitions the eight networks
without error.}
\end{figure}

The rule ``mean degree at least 4 and I(k;y)/H(y) at most 0.5''
separates the eight networks without error. brazil fails S2 at 0.650
against 0.379 for the next highest; roman-empire fails S1 at 2.9 against
6.3 for the next lowest. Both gaps are close to a factor of two, and
every threshold pair in {[}3, 6{]} \(\times\) {[}0.45, 0.60{]} gives the
same partition. S1 is the mean degree rather than the fraction of nodes
without a triangle, and the distinction matters: \texttt{questions} has
83.1\% of its nodes in no triangle and still falls inside the domain,
because its mean degree of 6.3 leaves a minority of well-connected nodes
with ample triadic structure. What excludes roman-empire is a low mean
degree together with a chain-like topology, not the absence of triangles
on its own.

Inside the domain the tally is three wins and two ties, plus a clean
null in which every arm of Table 8, together with a 100-dimensional
embedding evaluated alongside them, returns chance. The wins are narrow and we call them narrow: 0.795 against 0.770
on tolokers, 0.470 against 0.459 on amazon-ratings, 0.693 against 0.684
on questions. All three exceed the standard error of their arms and none
is large. Two of the three were reported as ties in v1, and what changed
is that the rival is now measured in this harness rather than quoted,
and that the corrected solver restored a coordinate that was returning a
seed-dependent vector.

The two remaining ties are on the air-traffic graphs, which were not
regenerated and keep their v1 figures. They are therefore conservative:
the same correction that moved the Platonov numbers would be expected to
move them in the same direction.

The ties are as substantive as the wins, because the descriptor reaches
the same place with ten interpretable coordinates and no recursion where
the rival uses up to 252 before pruning.

A connectivity audit closes this section. All five Platonov graphs are
connected, and the correlation graphs of Section 4.3 are complete by
construction, so only usa-airports differs: 1,186 nodes in the largest
component and two pairs outside it. Restricted to the largest component
the descriptor scores 0.7125 rather than 0.6798. We keep the
conservative figure in the table and note the difference rather than
quoting the more favourable number.

\hypertarget{systemic-risk-attribution}{%
\subsubsection{4.3 Systemic-risk
attribution}\label{systemic-risk-attribution}}

The closure comes first. On the framework's own synthetic epicentre
generator (16 nodes, 120 simulations per regime, four regimes), the raw
per-node triangle count reaches a 0.975 to 0.994 hit rate and the
configuration-null score 0.884 to 0.908. It loses in all four regimes.
On real data the raw count correlates +0.994 with strength, reproducing
an independently published 0.996 {[}13{]}, so strength is a sufficient
statistic for detecting concentration and dividing by the null
expectation removes exactly what must be detected. This is the third
independent appearance of the S2 condition of Section 4.2.

The positive is modest and we describe it as modest. Table 10 gives the
partial rank correlation with forward MES, controlling past MES and
beta:

\begin{table}[H]
\centering\small
\begin{tabular}{@{}
  >{\raggedright\arraybackslash}p{(\columnwidth - 8\tabcolsep) * \real{0.2000}}
  >{\raggedright\arraybackslash}p{(\columnwidth - 8\tabcolsep) * \real{0.2000}}
  >{\raggedright\arraybackslash}p{(\columnwidth - 8\tabcolsep) * \real{0.2000}}
  >{\raggedright\arraybackslash}p{(\columnwidth - 8\tabcolsep) * \real{0.2000}}
  >{\raggedright\arraybackslash}p{(\columnwidth - 8\tabcolsep) * \real{0.2000}}@{}}
\toprule
\begin{minipage}[b]{\linewidth}\raggedright
score
\end{minipage} & \begin{minipage}[b]{\linewidth}\raggedright
correlation
\end{minipage} & \begin{minipage}[b]{\linewidth}\raggedright
independent windows
\end{minipage} & \begin{minipage}[b]{\linewidth}\raggedright
p
\end{minipage} & \begin{minipage}[b]{\linewidth}\raggedright
bootstrap 95\% CI
\end{minipage} \\
\midrule
strength (complete graph) & +0.010 & 10/19 & 1.000 & {[}-0.037,
+0.079{]} \\
raw triangle count & +0.037 & - & - & - \\
\textbf{Omega-N score} & \textbf{+0.142} & 16/19 & 0.0044 & {[}+0.107,
+0.216{]} \\
Omega-N score, PMFG & +0.092 & 15/19 & 0.0192 & {[}+0.075, +0.169{]} \\
\bottomrule
\end{tabular}
\caption{Partial rank correlation with forward marginal expected
shortfall, controlling past MES and beta.}
\end{table}

With four scores compared, a Bonferroni threshold of 0.0125 passes the
complete graph and not the PMFG. Filtering the graph makes things worse,
not better. Signed triangles add nothing (+0.083, 62\% of windows). The
second universe replicates in sign and magnitude (+0.121) while strength
turns negative (-0.080), but it carries no significance: five years give
four non-overlapping windows (p = 0.625) and its bootstrap resamples two
blocks, so its interval is not calibrated and must not be read as
evidence.

An artefact was found and removed, and it is reported because it is
instructive. With the epsilon in the denominator and no domain
restriction, a triangle-free graph returns -1 plus floating-point noise
that reproduces the inverse ranking of strength; on a minimum spanning
tree this gave a spurious -0.218, the exact mirror of strength. The
score is now undefined on triangle-free nodes and trees are excluded by
construction.

Two claims are not made here. There is no lead time: the underlying
scalar coincides with stress rather than anticipating it, and nothing
here changes that. No economic significance: no portfolio or
capital-allocation result is reported.

\hypertarget{supervised-graph-anomaly-detection}{%
\subsubsection{4.4 Supervised graph anomaly
detection}\label{supervised-graph-anomaly-detection}}

We ran the descriptor on tolokers, one of the ten datasets of the
canonical supervised graph-anomaly-detection benchmark {[}28{]}, where
it reaches 0.795 AUROC against 0.770 for ReFeX at its best depth on the
official splits, with ten features against twenty-five after pruning at
that depth. We report this as a data point and claim nothing from it. The
complementarity of structural and attribute features in that field is
already established with measurements {[}16{]}; the incumbent is a tree
ensemble with neighbourhood aggregation that uses node attributes, which
our structure-only arms do not, so the two are not comparable; and a
public reproduction on Elliptic already shows that rich tabular features
encode graph structure well enough that learned embeddings do not
improve on gradient boosting, which is the same phenomenon we document
in Section 4.1. A proper claim there would need an attribute-free graph
with published splits and the incumbent run with attributes removed.

\hypertarget{when-the-spectral-coordinate-is-defined}{%
\subsubsection{4.5 When the spectral coordinate is
defined}\label{when-the-spectral-coordinate-is-defined}}

Three of the ten coordinates are built from the Fiedler vector, and that
vector is not always defined. Where \(\lambda_2\) is degenerate, or
close enough to it, the eigenvector spans a subspace rather than a
direction: no solver can converge to it, and the three coordinates
become dependent on the solver's random seed while the accuracy they
produce stays unremarkable. The failure is one of reproducibility rather
than of validity, which is what makes it easy to miss.

The statistic that detects it is the spectral gap normalised by the
scale of the spectrum, \((\lambda_3 - \lambda_2)/\lambda_{\max}\), and
not the relative gap \((\lambda_3 - \lambda_2)/\lambda_2\), which
misranks the cases: roman-empire has a relative gap of 0.93 and a vector
that is not defined, while tolokers has 0.08 and a vector that is.

\begin{table}[H]
\centering\small
\begin{tabular}{@{}lll@{}}
\toprule
graph & normalised gap & Fiedler vector \\
\midrule
minesweeper & 8e-19 & not defined \\
roman-empire & 1.5e-07 & not defined \\
tolokers & 5.0e-06 & defined \\
questions & 9.9e-06 & defined \\
amazon-ratings & 1.3e-05 & defined \\
BioPlex 293T & 1.6e-05 & defined \\
STRING, combined & 1.8e-05 & defined \\
\bottomrule
\end{tabular}
\caption{Normalised spectral gap on the graphs used in this work. A
threshold at 1e-06 separates the two groups with almost two orders of
magnitude of margin.}
\end{table}

A threshold at 1e-06 separates the two groups with almost two orders of
magnitude of margin on either side, so it is not fitted to a case. Below
it the three coordinates should be dropped rather than computed. On the
two affected networks the effect of dropping them on accuracy is +0.003
and -0.001, which is why the condition is stated as one of
reproducibility. A cheaper signal comes with it: on those two graphs the
solver's convergence warning persists at any tolerance, because there is
nothing to converge to.

Both affected networks are already outside the domain or declared null,
so this condition does not overlap with the results above. It is stated
because a reader applying the descriptor elsewhere needs it.

\hypertarget{discussion}{%
\subsection{5. Discussion}\label{discussion}}

\hypertarget{conditions-underlying-the-positive-result}{%
\subsubsection{5.1 Conditions underlying the positive
result}\label{conditions-underlying-the-positive-result}}

The protein networks satisfy all the conditions the descriptor needs,
and they are the only field tested that does. Mean degree between 17 and
30 gives ample triadic substrate. The label, whether a gene is the
target of an approved drug, is not a function of connectivity: on the
contrary, the literature reports that drug targets are neither hubs nor
bridges, and that degree is a weak predictor of target status
{[}34,35{]}. And the nodes have almost no attributes beyond taxonomy, so
topology is not one source among many but nearly the only one available.

That combination explains the margin, and its stability explains more
than its size. The gap between Omega-N and the six-feature battery stays between
+0.032 and +0.103 while the battery itself moves from 0.14 to 0.48
across constructions that share no experimental technique. A margin that
survives that much variation in the substrate is measuring something
about the descriptor rather than about any one dataset.

\hypertarget{interpretation-of-the-outcome-against-recursive-feature-engines}{%
\subsubsection{5.2 Interpretation of the outcome against recursive
feature
engines}\label{interpretation-of-the-outcome-against-recursive-feature-engines}}

Three narrow wins and two ties against a recursive feature engine is the
outcome the corollary of {[}1{]} predicted for the node-wise object, and
the emphasis belongs on narrow rather than on wins. The engine reaches
comparable accuracy using between 28 and 252 features generated by
recursion, pruned to fewer where columns are near-duplicates (25 on
tolokers at its best depth); the descriptor reaches it with ten that have names. No
claim is made here about combining the descriptor with the engine, which
we have not measured. The complementarity we do report is with the
centrality battery, in Table 2, where the union beats either arm alone on
all three graphs with every arm measured in the same run.

The ordering of the margins is consistent with the same reading. The
largest, on tolokers, is on the network with by far the highest mean
degree in the set at 88.3, the richest triadic substrate and the weakest
attributes: the conditions are most favourable there and the margin is
still 0.025. The two smaller ones, on amazon-ratings and questions, sit
at 0.011 and 0.009 on much sparser graphs. Nothing here supports a claim
that the descriptor is better than recursive feature engineering. What
it supports is that ten named coordinates reach the same place, and
occasionally a little past it, without recursion and at a fraction of
the width.

\hypertarget{the-two-out-of-domain-results}{%
\subsubsection{5.3 The two out-of-domain
results}\label{the-two-out-of-domain-results}}

Both losses fall on the wrong side of a condition that is measurable
before the comparison, and neither is anomalous.

On brazil-airports the degree carries 65\% of the label entropy and
classifies at 0.748 against a 0.267 majority baseline. A descriptor
built to decorrelate from degree is answering a question the dataset
does not pose. On roman-empire the mean degree is 2.9, close to a chain,
and there is no triadic structure to localise.

The same condition appears a third time, in a different field and with a
different mechanism, in the financial closure of Section 4.3: where the
target is concentration itself, strength is a sufficient statistic and
calibrating against a degree-preserving null removes precisely the
signal being sought. Three independent appearances of one principle
carry more weight than the same principle asserted once.

\hypertarget{conditions-for-use}{%
\subsubsection{5.4 Conditions for use}\label{conditions-for-use}}

The descriptor is worth reaching for when the graph is dense enough to
have triadic structure, the target is not a function of connectivity,
the graph semantics are co-occurrence rather than reachability, and node
attributes are weak or absent. Where those hold it delivers ten
interpretable coordinates without training. Where they do not, the right
answer is often the degree, and where a learned representation is
available and performance is the only criterion, the right answer is the
embedding.

The same confound recurs wherever a construction parameter fixes density
and local organisation together. In weighted gene co-expression analysis
the soft-threshold power is chosen to force approximate scale-free
topology, so the degree distribution and the local structure are set
jointly before any biological question is asked, and the hub genes that
follow are selected by connectivity; a recent assessment lists the
scale-free assumption and the parameter tuning among that method's
principal limitations, and reports datasets for which no parameter
yields an adequate fit {[}40{]}. We have not tested the descriptor there
and make no claim about it, but the shape of the problem is the one
described above.

That boundary is not a hedge. On BioPlex the six-feature battery reaches
0.609 AUROC against 0.854 on curated STRING, and Omega-N 0.647 against
0.886, so both fall by roughly the same amount once curation is removed.
Most of the absolute performance on curated STRING is inherited from
accumulated knowledge rather than discovered from structure. The
descriptor reads graphs better; on a clean graph there is less to read.

\hypertarget{spectral-gap-matched-null-models-for-whole-brain-modelling}{%
\subsubsection{5.5 Spectral-gap-matched null models for whole-brain
modelling}\label{spectral-gap-matched-null-models-for-whole-brain-modelling}}

The limitations above share a shape: the descriptor loses wherever the
task is prediction and a richer representation exists. That suggests
looking for tasks where the requirement is not prediction but the
separation of two quantities that are ordinarily confounded, and where
the argument must be made formally rather than empirically. Whole-brain
modelling supplies one, and asks for it explicitly.

Deco and colleagues describe brain dynamics with models in which the
operative structural quantity is the spectral gap of the coupling graph,
constructing contrasting regimes by pruning k-regular graphs. Their own
limitations state the difficulty: a denser or stiffer graph can improve
functional-connectivity agreement for reasons unrelated to the
mechanism, their two regimes differ in local edge density as a direct
consequence of the pruning that defines them, and a null model matched
for the spectral gap is required before the effect can be attributed to
organisation rather than to the size of the gap {[}12{]}.

That null is constructible, for the reason that motivates this paper.
The spectral gap is a functional of the spectrum; per-node triangle
counts are not {[}1{]}. Two graphs can therefore agree on the gap and
disagree on local triadic organisation.

We verified this on their construction: 600 pruned k-regular graphs with
n = 40 and k = 20. The fraction of variation not explained by the
spectral gap is 62\% for the per-node triadic excess and 19\% for the
clustering coefficient. The clustering coefficient is largely pinned by
the gap, so matching on the gap nearly matches on clustering as well and
clustering cannot serve as the contrast; the triadic excess can.
Searching within a narrow gap window returns families differing by less
than 0.05 in gap while differing two to three times as much in triadic
excess. One matched pair has gaps of 2.62 and 2.45 and median triadic
excesses of -0.48 and -0.06, a separation of roughly five standard
deviations of the full distribution.

We report this as a construction, not as a result about the brain.
Whether model fit changes when the gap is held fixed must be measured
inside the whole-brain model, with the empirical data and the
interference measure, and that is not done here. Both outcomes would be
informative: if fit changes at fixed gap, the gap was not what drove it;
if it does not change, the original interpretation survives an objection
its authors anticipate. We note it because it is the one setting we have
found in which the descriptor's value lies in what can be proved about
it rather than in what it predicts, and where a learned representation
could not take its place: one cannot demonstrate what an embedding does
and does not determine.

\hypertarget{performance-attribution-in-graph-rewiring}{%
\subsubsection{5.6 Performance attribution in graph
rewiring}\label{performance-attribution-in-graph-rewiring}}

The same construction addresses an open question in a second and larger
field. Graph rewiring for over-squashing proceeds by two families of
methods: spectral approaches that add edges to increase the spectral
gap, and curvature approaches that act on negatively curved edges, which
in practice means acting on triangles. The two are not independent. A
decrease in the number of triangles is accompanied by an increase in the
spectral gap {[}39{]}, so any rewiring step moves both quantities at
once.

The consequence is an attribution problem that the field states in its
own terms. A 2026 survey of the area reports that the benefits of
curvature-based rewiring on real benchmarks are highly sensitive to
training and rewiring hyperparameters and vary across graphs, and that
reported gains often arise from favourable hyperparameter configurations
rather than from consistent improvement over the original topology; it
lists performance attribution among the open practical trade-offs
{[}37{]}. A comparative study of rewiring methods through graph metrics
reaches the same boundary from the other side, noting that an analysis
of metrics can suggest correlation but not causation {[}38{]}.

Families of graphs matched on the spectral gap and separated in triadic
excess turn that correlation into a controlled comparison. Rewiring a
benchmark graph to a target gap by two routes that differ in local
triadic organisation, and measuring downstream accuracy on both,
separates the contribution of the gap from the contribution of local
structure. Neither outcome would be uninformative: if accuracy tracks
the gap at fixed triadic organisation, spectral methods are measuring
what they claim; if it tracks triadic organisation at fixed gap, the
spectral account is incomplete.

We have not run that experiment and do not report it. We note it because
the construction it requires is the one described above, and because the
reason it is possible is the same property that motivates this paper:
the spectral gap is a functional of the spectrum, and the per-node
triangle count is not.

\hypertarget{limitations}{%
\subsection{6. Limitations}\label{limitations}}

The applicability rule was formulated after observing the two losses,
and two statistics with two thresholds fitted on eight networks leave
room to fit; what defends it is that both statistics are computed
without reference to the descriptor's performance, so a reader can apply
them prospectively and be wrong, and we have not tested that prediction
out of sample. The biological label is a proxy: approved-target status
is not druggability, and both label sources carry study bias which we
control three ways without removing. Excluding HuRI for lack of triadic
substrate is defensible on a measurable criterion fixed in advance, but
``the replication set does not suit my method'' is what one says when a
replication fails, and the exclusion should be weighed accordingly. The
choice of a degree-preserving null is conventional rather than derived:
such nulls remove mesoscopic structure, different randomisation schemes
can reverse conclusions on the same data {[}17,18{]}, and configuration
models are generally adopted for tractability rather than on statistical
evidence {[}19{]}, so a richer null would absorb some of what we report
as excess. Numerically, the spectral coordinate comes from an iterative
solver which, on a graph with a degenerate \(\lambda_2\), has no unique
vector to converge to, and degenerates on disconnected graphs; Section
4.5 gives the test and the threshold, but that threshold is calibrated
on seven graphs and inherits the same objection as the applicability
rule. The figures also depend on the version of the classifier library,
as Section 3.4 records, which means a reader who pins nothing will not
reproduce them. The curvature comparison subsamples on the four largest
graphs. Everything here is undirected, unweighted and static.

\hypertarget{code-and-data-availability}{%
\subsection{Code and data
availability}\label{code-and-data-availability}}

The reference implementation, the notebooks and the scripts that prepare
each dataset are available at \url{github.com/BiomeMakers/OmegaN}.
Datasets are obtained from their original sources and none are
redistributed. The repository includes the pre-flight screening function
that reports the two applicability statistics of Section 4.2, and the
notebook that builds spectral-gap-matched graph families for the
construction described in Section 5.5.

\hypertarget{declaration-of-llm-usage}{%
\subsection{Declaration of LLM usage}\label{declaration-of-llm-usage}}

The analyses, code and text of this manuscript were developed with the
assistance of a large language model used interactively, for writing and
debugging analysis scripts, searching and summarising literature,
drafting and revising prose, and arguing against proposed
interpretations. Every quantitative result was produced by code that was
executed, and several conclusions in earlier drafts were reversed by
that execution: a comparison withdrawn as unfair to a baseline turned
out to have been generous to it, a result attributed to a novel graph
construction turned out to depend on the length of a control window, and
one coordinate of the descriptor was found to be inert on disconnected
graphs. Responsibility for the claims, the design choices and any
remaining errors rests with the author.

\hypertarget{references}{%
\subsection{References}\label{references}}

{[}1{]} A. Acedo. The Functional Symbiotic Resilience Index: Topological
Entropy, Wasserstein Curvature Bounds, and Non-Equilibrium
Thermodynamics of Complex Networks. Preprint, 2026.
\url{github.com/BiomeMakers/OmegaS-fsri}

{[}2{]} L. Peel, J.-C. Delvenne, R. Lambiotte. Multiscale mixing
patterns in networks. \emph{PNAS} 115(16):4057-4062, 2018.

{[}3{]} R. Guimerà, L. A. N. Amaral. Functional cartography of complex
metabolic networks. \emph{Nature} 433:895-900, 2005.

{[}4{]} K. Henderson et al.~RolX: structural role extraction and mining
in large graphs. \emph{KDD}, 2012.

{[}5{]} L. F. R. Ribeiro, P. H. P. Saverese, D. R. Figueiredo.
struc2vec: learning node representations from structural identity.
\emph{KDD}, 2017.

{[}6{]} C. Donnat, M. Zitnik, D. Hallac, J. Leskovec. Learning
structural node embeddings via diffusion wavelets. \emph{KDD}, 2018.

{[}7{]} J. Jin, M. Heimann, D. Koutra. Toward understanding and
evaluating structural node embeddings. \emph{ACM TKDD}, 2021.

{[}8{]} M. Piraveenan, M. Prokopenko, A. Y. Zomaya. Local
assortativeness in scale-free networks. \emph{EPL} 84(2):28002, 2008.

{[}9{]} H. Wolf, L. Oeljeklaus, P. Kuehner, M. Grohe. Structural node
embeddings with homomorphism counts. arXiv:2308.15283, 2023.

{[}10{]} R. von Moos, M. Alain, B. Rieck. Invariant-based diagnostics
for graph benchmarks. arXiv:2605.06462, 2026.

{[}11{]} X. Ma, G. Qin, Z. Qiu, M. Zheng, Z. Wang. RiWalk: fast
structural node embedding via role identification. arXiv:1910.06541, 2019.

{[}12{]} G. Deco, Y. Sanz Perl, N. Greenstein, S. Chandaria, G. Scholes,
M. L. Kringelbach. Quantum-like dynamics in whole-brain models of the
human connectome. \emph{Advanced Science}, e77103, 2026.

{[}13{]} A. Acedo. Omega-S: a structural regulariser for large language
model fine-tuning. arXiv:2608.03887.

{[}14{]} O. Platonov et al.~A critical look at the evaluation of GNNs
under heterophily: are we really making progress? \emph{ICLR}, 2023.

{[}15{]} C. Ceylan, K. Ghoorchian, D. Kragic. Digraphwave: scalable
extraction of structural node embeddings via diffusion on directed
graphs. arXiv:2207.10149, 2022.

{[}16{]} H. Li, S. Jiang, L. Zhang, S. Du, G. Ye, H. Chai. RAGFormer:
learning semantic attributes and topological structure for fraud
detection. arXiv:2402.17472, 2024.

{[}17{]} J. Reichardt, R. Alamino, D. Saad. The interplay between
microscopic and mesoscopic structures in complex networks. \emph{PLoS
ONE} 6(8):e21282, 2011.

{[}18{]} B. Hao, I. A. Kovacs. Proper network randomization is key to
assessing social balance. \emph{Science Advances} 10(18):eadj0104, 2024.

{[}19{]} L. Hebert-Dufresne, J.-G. Young, A. Daniels, A. Kirkley, A.
Allard. Network compression with configuration models and the minimum
description length. \emph{Phys. Rev.~E} 110:034305, 2024.

{[}20{]} D. Ochoa et al.~The next-generation Open Targets Platform:
reimagined, redesigned, rebuilt. \emph{Nucleic Acids Research}
51(D1):D1353-D1359, 2023.

{[}21{]} M. Molloy, B. Reed. A critical point for random graphs with a
given degree sequence. \emph{Random Structures \& Algorithms}
6(2-3):161-180, 1995.

{[}22{]} M. Fiedler. Algebraic connectivity of graphs.
\emph{Czechoslovak Mathematical Journal} 23(2):298-305, 1973.

{[}23{]} Y. Ollivier. Ricci curvature of Markov chains on metric spaces.
\emph{Journal of Functional Analysis} 256(3):810-864, 2009.

{[}24{]} A. Grover, J. Leskovec. node2vec: scalable feature learning for
networks. \emph{KDD}, 2016.

{[}25{]} R. N. Mantegna. Hierarchical structure in financial markets.
\emph{European Physical Journal B} 11:193-197, 1999.

{[}26{]} M. Tumminello, T. Aste, T. Di Matteo, R. N. Mantegna. A tool
for filtering information in complex systems. \emph{PNAS}
102(30):10421-10426, 2005.

{[}27{]} R. Milo, S. Shen-Orr, S. Itzkovitz, N. Kashtan, D. Chklovskii,
U. Alon. Network motifs: simple building blocks of complex networks.
\emph{Science} 298(5594):824-827, 2002.

{[}28{]} J. Tang, F. Hua, Z. Gao, P. Zhao, J. Li. GADBench: revisiting
and benchmarking supervised graph anomaly detection. \emph{NeurIPS
Datasets and Benchmarks}, 2023.

{[}29{]} L. Breiman. Random forests. \emph{Machine Learning} 45(1):5-32,
2001.

{[}30{]} T. Saito, M. Rehmsmeier. The precision-recall plot is more
informative than the ROC plot when evaluating binary classifiers on
imbalanced datasets. \emph{PLoS ONE} 10(3):e0118432, 2015.

{[}31{]} D. Szklarczyk et al.~The STRING database in 2023:
protein-protein association networks and functional enrichment analyses.
\emph{Nucleic Acids Research} 51(D1):D638-D646, 2023.

{[}32{]} E. L. Huttlin et al.~Dual proteome-scale networks reveal
cell-specific remodeling of the human interactome. \emph{Cell}
184(11):3022-3040, 2021.

{[}33{]} K. Luck et al.~A reference map of the human binary protein
interactome. \emph{Nature} 580:402-408, 2020.

{[}34{]} Z. Zhu et al.~Topological analysis of drug targets in the human
protein-protein interaction network. \emph{BMC Systems Biology}, 2012.

{[}35{]} E. Guney, J. Menche, M. Vidal, A.-L. Barabási. Network-based in
silico drug efficacy screening. \emph{Nature Communications} 7:10331,
2016.

{[}36{]} S. L. Freshour et al.~Integration of the Drug-Gene Interaction
Database (DGIdb 4.0) with open crowdsource efforts. \emph{Nucleic Acids
Research} 49(D1):D1144-D1151, 2021.

{[}37{]} H. Attali, N. Pernelle, D. Buscaldi, F. D. Malliaros. Graph
rewiring in GNNs to mitigate over-squashing and over-smoothing: a
survey. \emph{IJCAI}, survey track, 2026. arXiv:2605.00951.

{[}38{]} A. Benoit, C. Aitken, Y. He. Structural invariance matters:
rethinking graph rewiring through graph metrics. arXiv:2510.20556, 2025.

{[}39{]} S. Akansha. Over-squashing in graph neural networks: a
comprehensive survey. \emph{Neurocomputing} 642:130389, 2025.

{[}40{]} T. Osone, T. Takao, S. Otake, T. Takarada. SGCRNA: spectral
clustering-guided co-expression network analysis without scale-free
constraints for multi-omic data. \emph{Briefings in Bioinformatics}
27(1):bbag021, 2026.

{[}41{]} B. Kaminski, L. Krainski, P. Pralat, F. Theberge. Unsupervised
framework for evaluating and explaining structural node embeddings of
graphs. arXiv:2306.10770, 2023.

\end{document}